\documentclass[aps,prl,twocolumn,showpacs]{revtex4}

\usepackage{amssymb,amsmath}
\usepackage{epsfig}
\usepackage{bbm}
\usepackage{dsfont}

\newlength{\upit}\upit=0.1truein

\newcommand{\ltappr}{{{\lower4pt\hbox{$<$} } \atop \widetilde{ \ \ \ }}}
\newlength{\bxwidth}\bxwidth=1.5 truein

\newcommand{\dg}{^{\dagger }}

\newcommand{\rarrow}{\rightarrow}
\newcommand{\beg}{\begin{equation}}
\newcommand{\en}{\end{equation}}
\newcommand{\bq}{\mathbf q}

\newlength{\figwidth}
\newlength{\shift}
\usepackage{color,ulem}

\newcommand \bea {\begin{eqnarray} }
\newcommand \eea {\end{eqnarray}}
\newcommand{\bk}{{\bf{k}}}
\newcommand{\bx}{{\bf{x}}}
\newcommand{\by}{{\bf{y}}}
\newcommand{\bz}{{\bf{z}}}
\newcommand{\hx}{\hat {\bf{x}}}
\newcommand{\hy}{\hat {\bf{y}}}
\newcommand{\hz}{\hat {\bf{z}}}
\newcommand{\bR}{{\bf{R}}}
\newcommand{\pmat}[1]{\begin{pmatrix}#1\end{pmatrix}}
\newcommand\smb{SmB$_{6}$}

\begin{document}

\title{Cubic Topological Kondo Insulators}

\author{Victor Alexandrov$^{1}$, Maxim Dzero$^{2}$ and Piers Coleman$^{1,3}$}

\affiliation{$^{1}$Center for Materials Theory, Department of Physics and Astronomy,
Rutgers University, Piscataway, NJ 08854-8019, USA}
\affiliation{$^{2}$ Department of Physics, Kent State University,
Kent, OH 44242, USA}
\affiliation{$^{3}$ Department of Physics, Royal Holloway, University
of London, Egham, Surrey TW20 0EX, UK.}
\date{\today}
\begin{abstract}
Current theories of Kondo insulators employ the interaction of
conduction electrons with localized Kramers doublets originating
from a tetragonal crystalline environment, yet all
Kondo insulators are cubic.  Here we develop a theory of
cubic topological Kondo insulators involving the interaction of $\Gamma_{8}$ spin
quartets with a conduction sea.  The spin quartets greatly increase
the potential for strong topological insulators, entirely eliminating
the weak-topological phases from the diagram.  We show that the
relevant topological behavior in cubic Kondo insulators can only
reside at the lower symmetry X or M points in the Brillouin zone, leading to three Dirac
cones with heavy quasiparticles.
\end{abstract}

%
\maketitle
%


Our classical understanding of order in matter is built around Landau's
concept of an order parameter. The past few years have seen
a profound growth of interest in  topological phases of matter,
epitomized by the quantum Hall effect and topological band insulators,
in which the underlying order derives from the non-trivial connectedness of the
quantum wave-function, often driven by the presence of strong
spin-orbit coupling \cite{Fu2007,Moore2007,Hsieh2008,Xia2009,exp1,exp2,exp3,Hasan2010,Qi2010}.

One of the interesting new entries to the world of topological
insulators, is the class of heavy fermion, or ``Kondo
insulators'' \cite{KIReviews1,KIReviews2,Dzero2010,Takimoto2011,Dzero2012,Tran2012,Dai2012}.
The strong-spin orbit coupling and highly renormalized narrow bands in these intermetallic materials inspired
the prediction \cite{Dzero2010} that a subset of the family of Kondo insulators will be
Z$_{2}$ topological insulators. In particular, the oldest known Kondo
insulator \smb  \cite{Geballe1969} with marked mixed valence character, was
identified as a particularly promising candidate for a strong
topological insulator (STI): a conclusion that has since also been
supported by band-theory calculations \cite{Takimoto2011,Dai2012}. Recent experiments \cite{exp1smb6,exp2smb6,exp3smb6}
on \smb{} have confirmed the presence of robust conducting surfaces,
large bulk resistivity and a chemical potential that clearly lies in
the gap providing strong support for the initial prediction.

However, despite these developments, there are still many
aspects  of the physics in these materials that are poorly understood.
One of the simplifying assumptions of the original theory \cite{Dzero2010} was to
treat the $f$-states as Kramer's doublets in a tetragonal
environment.  In fact,  the tetragonal theory
predicts that strong topological insulating behavior requires large
deviations from integral valence,
while in practice Kondo insulators are much closer to integral valence \cite{KIReviews2}.
Moreover, all known Kondo insulators have cubic
symmetry, and this higher symmetry
appears to play a vital role, for
all apparent ``Kondo insulators'' of lower symmetry, such as
CeNiSn\cite{cenisn} or CeRu$_{4}$Sn$_{6}$\cite{cerusn6} have proven, on improving sample quality, to be
semi-metals. One of the important effects of high symmetry is the
stablization of magnetic f-quartets. Moreover,
Raman\cite{Sarrao1997} experiments and various band-theory studies\cite{Yanase,Antonov} that it is the Kondo screening of the magnetic quartets that gives rise to the emergence of the insulating state.

\begin{figure}[h]
\includegraphics[scale=0.08,angle=0]{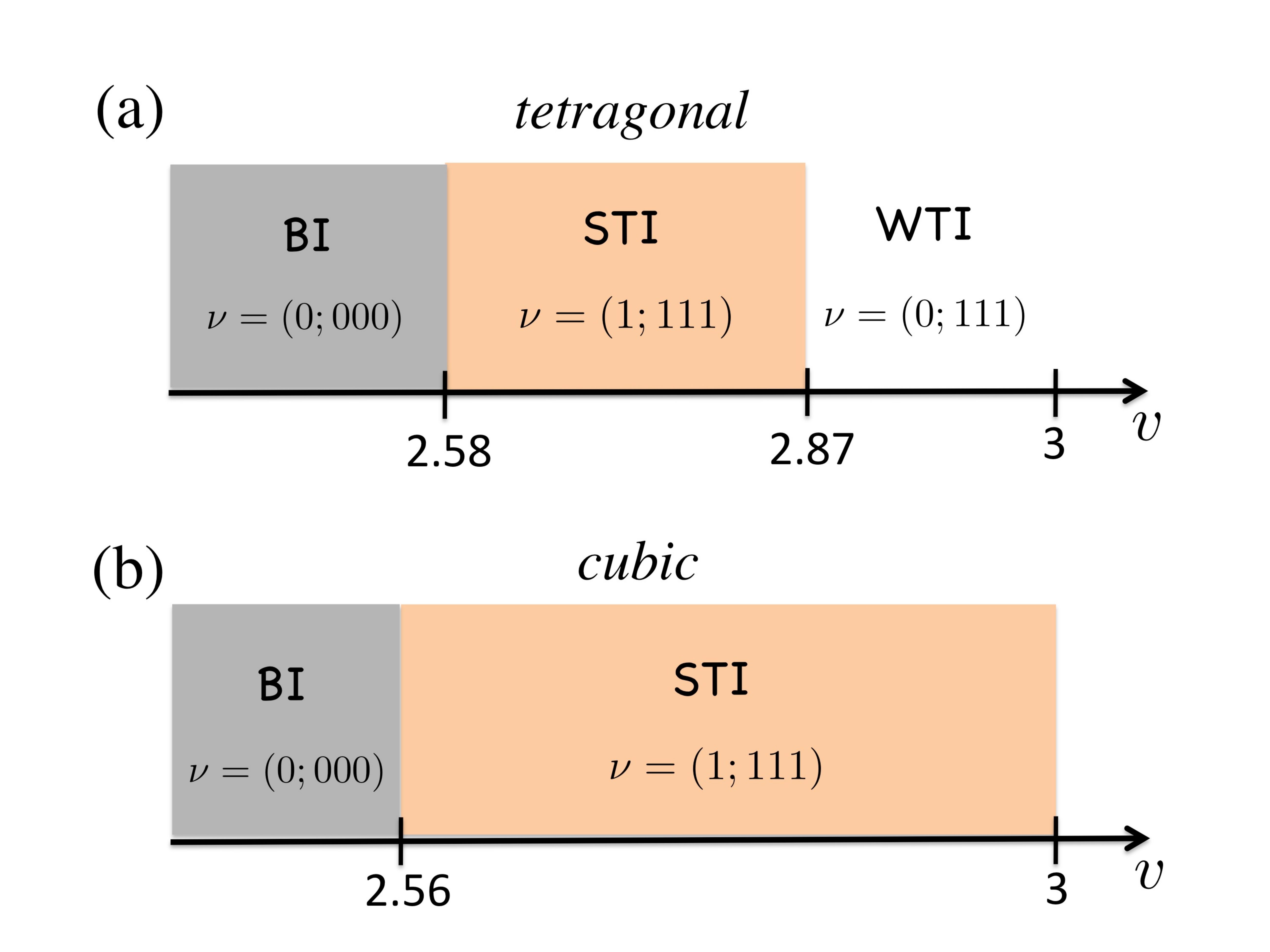}
\caption{
Contrasting the phase diagram of tetragonal
\cite{DzeroLargeN} and cubic topological Kondo insulators. Cubic
symmetry extends the STI phase into the
Kondo limit. For \smb{} $v=3-n_{f}$ gives the valence of the Sm ion, while $n_f$
measures the number of $f$-holes in the filled $4f^{6}$ state, so
that $n_f=1$ corresponds to the $4f^{5}$ configuration.
}
 \label{Fig1}
\end{figure}

Motivated by this observation, here we formulate a theory of cubic
topological Kondo insulators, based on a lattice of magnetic quartets.  We
show that the presence of a spin-quartet greatly increases the possibility of strong
topological insulators while eliminating the
weak-topological insulators from the phase diagram, Fig. \ref{Fig1}.  We
predict  that the relevant topological behavior in simple cubic Kondo
insulators can only reside at the lower point group symmetry X and M points in the Brillouin zone (BZ), leading to a
three heavy Dirac cones at the surface. One of the additional consequences of the underlying Kondo physics,
is that the  coherence length of the surface states is  expected to be
very small, of order a lattice spacing.

While we outline our model of cubic Kondo insulators with a
particular focus on \smb{}, the methodology
generalizes to  other cubic Kondo insulators.
\smb{} has a simple cubic structure, with the B$_{6}$ clusters located
at the center of the unit cell, acting as spacers which mediate electron hopping between Sm sites.
Band-theory \cite{Antonov} and XPS studies
\cite{Arpes} show that the 4f orbitals hybridize with $d$-bands
which form electron pockets around the X points.  In a cubic environment, the
 $J=5/2$  orbitals split into a $\Gamma_{7}$ doublet and a $\Gamma_{8}$ quartet, while the fivefold degenerate
$d$-orbitals are split into double degenerate $e_g$ and triply degenerate $t_{2g}$ orbitals.
Band theory and Raman spectroscopy studies  \cite{Sarrao1997} indicate that the physics of the $4f$ orbitals is governed by valence fluctuations involving electrons of the $\Gamma_{8}$ quartet and the conduction $e_g$ states, $e^{-}+4f^{5}
(\Gamma_{8}^{(\alpha)})\rightleftharpoons4f^{6}$.
The $\Gamma_8^{(\alpha)}$ ($\alpha=1,2$) quartet consists of the following combination of orbitals: $|\Gamma_8^{(1)}\rangle=\sqrt{\frac{5}{6}}\left\vert\pm\frac{5}{2}\right\rangle+ \sqrt{\frac{1}{6}}\left\vert\mp\frac{3}{2}\right\rangle,~ |\Gamma_8^{(2)}\rangle=\left\vert\pm\frac{1}{2}\right\rangle$. This then leads to a simple physical picture in which the $\Gamma_{8}$ quartet of
$f$-states hybridizes with an $e_g$ {\it quartet} (Kramers plus orbital degeneracy) of $d$-states to form a Kondo insulator.

To gain insight into how the cubic topological Kondo
insulator emerges it is instructive to consider a simplified one-dimensional model consisting of a quartet of
conduction $d$-bands hybridized with a quartet of $f$-bands (Fig.~\ref{Fig2}a).
In one dimension there are two high symmetry points: $\Gamma$ ($k=0$) and X ($k=\pi$), where the
hybridization vanishes \cite{Dzero2010,Dzero2012,DzeroLargeN}).
Away from the zone center $\Gamma$, the $f-$ and $d-$ quartets
split into Kramers doublets.
The $Z_2$ topological invariant $\nu_{1D}$ is
then determined by the product of the parities $\nu_{1D}=\delta_{\Gamma}\delta_{X}$ of the occupied states at the $\Gamma$ and X points. However, the $f$-quartet  at the $\Gamma$ point is equivalent to
two Kramers doublets, which means that $\delta_{\Gamma}= (\pm
1)^{2}$
is always positive, so that $\nu_{1D}=\delta_X$ and a one-dimensional topological
insulator only  develops when the $f$ and $d$ bands invert at the X point.

\begin{figure}[h]
\includegraphics[width=0.45 \textwidth]{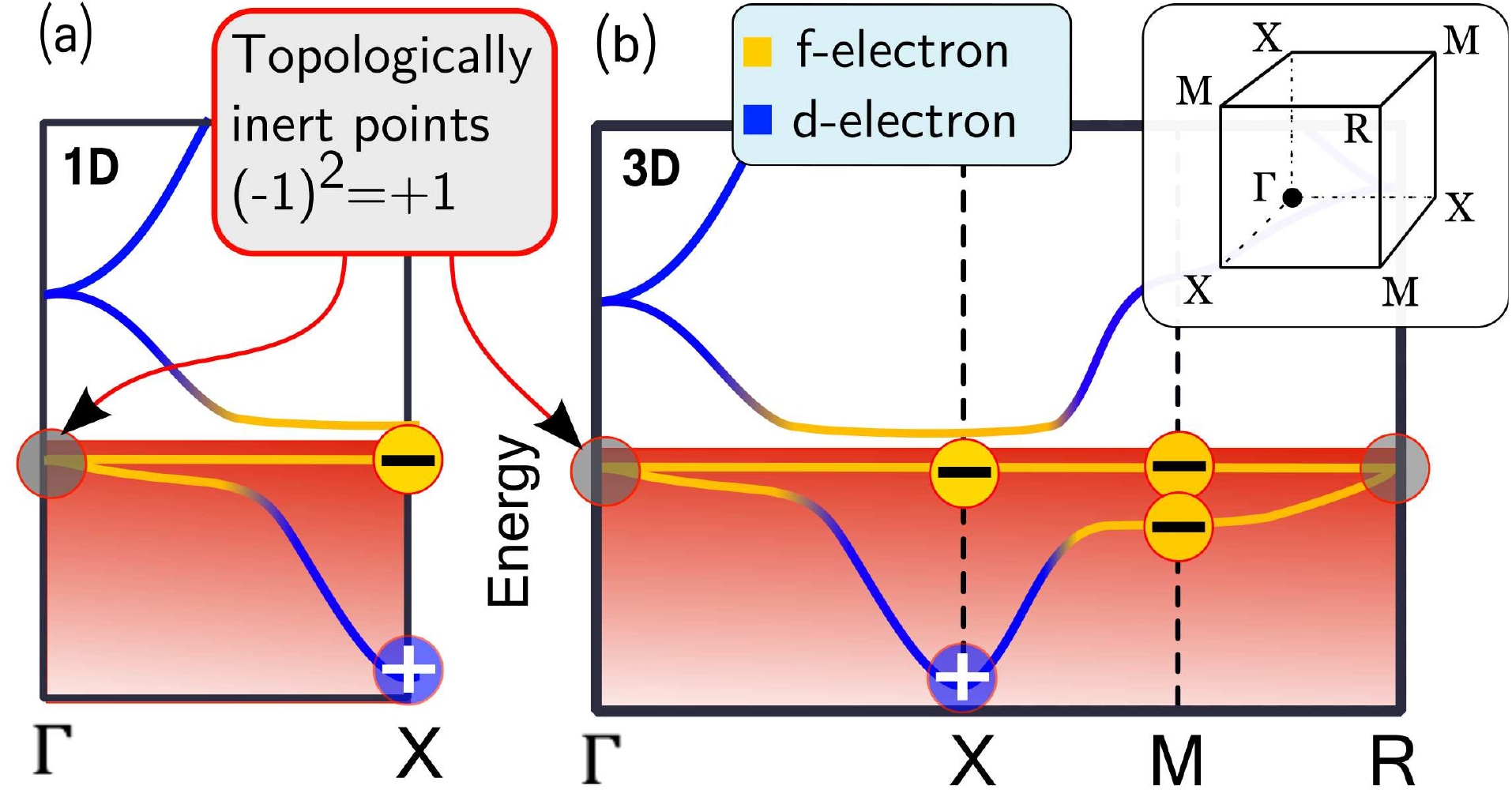}
\caption{
Schematic band structure illustrating (a) 1D Kondo insulator with
local cubic symmetry and (b) 3D cubic Kondo insulator. Hybridization
between a quartet of d-bands with a quartet of f-bands leads to a
Kondo insulator.
The fourfold degeneracy of the $f$- and $d$-bands at the
high symmetry $\Gamma$ and R points of the Brillouin zone guaranties that the 3D topological invariant
is determined by the band inversions at the X and M points only.
}
\label{Fig2}
\end{figure}

Generalizing this argument to three dimensions we see that
there are now four high symmetry points $\Gamma$, X, M and R.
The $f$-bands are fourfold degenerate at both $\Gamma$ and R
points which guarantees that
$\delta_{\Gamma}=\delta_{R}=+1$ (Fig.~\ref{Fig2}b). Therefore,  we see that the 3D topological invariant
is determined by band inversions at X or M points only, $\nu_{3D}=(\delta_X\delta_M)^3=\delta_{X}\delta_{M}$.
If there is a band inversion at the X point, we get $\nu_{3D} =
\delta_{X}\delta_{M} =-1$. In this way the cubic character of the Kondo insulator and, specifically, the fourfold
degeneracy of the $f$-orbital multiplet protects the formation of a
strong topological insulator.


We now formulate our model for cubic topological Kondo insulators.
At each site, the quartet of f and d -holes is described by an
orbital and spin index, denoted by the combination $\lambda\equiv
(a,\sigma )$ ($a=1,2$, $\sigma = \pm 1$). The fields are then given by the eight component spinor
\begin{equation}\label{}
\Psi_{j} = \left(\begin{matrix} d_{\lambda} (j)\cr X_{0\lambda } (j)
\end{matrix} \right)
\end{equation}
where $d_{\lambda} (j)$ destroys an d-hole at site j, while
$X_{0\lambda} (j)=\vert 4f^{6}\rangle \langle  4f^{5},\lambda\vert$ is the
the Hubbard operator that destroys an $f$-hole at site j.
The tight-binding Hamiltonian describing the
hybridized $f$-$d$ system is then
\begin{equation}\label{}
H = \sum_{i,j}\Psi\dg_{\lambda} (i)h_{\lambda\lambda'} (\bR_{i}-\bR_j) \Psi_{\lambda'} (j)
\end{equation}
in which the nearest hopping matrix has the structure
\begin{equation}\label{1d3d}
h (\bR)= \begin{pmatrix} h^{d} (\bR) & V (\bR)\cr V\dg (\bR)& h^{f} (\bR)
\end{pmatrix},
\end{equation}
where the diagonal elements describe hopping within the $d$- and $f$-
quartets while the off-diagonal parts describe the hybridization
between them, while $\bR \in (\pm \hx ,\pm \hy, \pm \hz)$ is the vector
linking nearest neighbors.  The various matrix elements simplify for
hopping along the z-axis, where they become orbitally and spin
diagonal:
\begin{equation}\label{}
h^{l}(\bz) = t^{l}\pmat{1 & \cr & \eta_{l}}, \qquad
V (\bz) =i  V \pmat{0 & \cr & \sigma_{z}}.
\end{equation}
where $l=d,f$ and $\eta_{l} $ is the ratio of orbital hopping elements.
In the above, the overlap between the $\Gamma_{8}^{(1)}$ orbitals,
which extend perpendicular to the z-axis is
neglected, since the hybridization is dominated by the overlap of the
the $\Gamma_{8}^{(2)}$ orbitals, which extend out along the z-axis.
The hopping matrix elements in the  $\bx $ and $\by$
directions are then obtained by rotations in orbital/spin space.
so that
$h (\bx )= U_{y}h (\bz)U_{y}\dg $ and $h (\by)= U_{-x}h (\bz
)U\dg_{-x}$ where $U_{y}$ and $U_{-x}$ denote 90$^{\circ }$ rotations
about the y and negative x axes, respectively.

The Fourier transformed hopping matrices $h (\bk )= \sum_{\bR} h (\bR)
e^{-i \bk \cdot\bR}$ can then be written in the compact form
\begin{equation}\label{Hdf3D}
h^{l} (\bk )= t^{l}\pmat{\phi_{1} (\bk )+ \eta_l\phi_{2} (\bk ) & (1-\eta_{l} )\phi_{3} (\bk )\cr
(1-\eta_{l} )\phi_{3} (\bk ) & \phi_{2} (\bk ) + \eta_{l} \phi_{1} (\bk )
}+ \epsilon^{l},
\end{equation}
where $l=d,f$. Here $\epsilon^{l}$ are the bare energies of the
isolated d and f-quartets, while
$\phi_{1} (\bk)= c_{x}+c_{y}+4 c_{z}$, $\phi_2(\bk) =
3(c_{x}+c_{y})$ and $\phi_{3} (\bk )= \sqrt{3}(c_{x}-c_{y})$ ($c_{\alpha}\cos
k_{\alpha}, \alpha=x,y,z$). The hybridization is given by
\begin{equation}\label{Hv3D}
V (\bk )=  \frac16
\pmat{
   3 (\bar\sigma_x  +i \bar\sigma_y  )& \sqrt3( \bar\sigma_x  -i \bar\sigma_y  ) \cr
   \sqrt3( \bar\sigma_x  -i \bar\sigma_y  ) &\bar \sigma_x + i\bar \sigma_y  + 4 \bar\sigma_z  \cr
}
\end{equation}
where we denote $\bar\sigma_\alpha = \sigma_\alpha \sin k_{\alpha}$.
Note how the hybridization
between the even parity d-states and odd-parity f-states is an odd
parity function of momentum $V (\bk )= - V (-\bk )$.

To analyze the properties of the Kondo insulator, we use a slave boson
formulation of the Hubbard operators, writing $X_{\lambda 0} (j) =
f\dg_{\lambda} (j)b_{j}$, where $f\dg_{\lambda}\vert 0 \rangle  \equiv \vert
4f^{5},\lambda\rangle $ creates an f-hole in the $\Gamma^{8}$ quartet
while   $b\dg\vert 0\rangle \equiv  \vert 4f^{6}\rangle $ denotes the
singlet filled $4f$ shell, subject to the constraint
$Q_{j}= b\dg_{j}b_{j}+\sum_{\lambda} f\dg_{j\lambda}f_{j\lambda}=1$ at each site.

\begin{figure}[h]
\includegraphics[scale=0.28,angle=0]{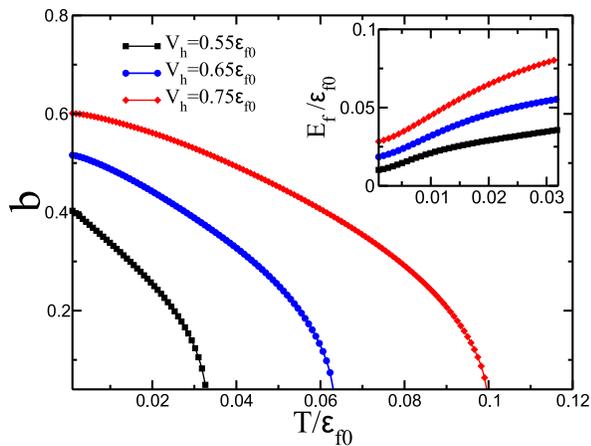}
\caption{Temperature dependence of the
hybridization gap parameter $b$ and the renormalized $f$-level
position (inset) for various values of the bare hybridization (see Supplementary Materials for more details). }
\label{Fig3}
\end{figure}

%
%
We now analyze the properties of the cubic Kondo insulator,
using a mean-field treatment of the slave boson field $b_{i}$,
replacing the
slave-boson operator $\hat{b}_i$ at each site by its expectation value:
$\langle \hat{b}_i \rangle =b$ so that the f- hopping and
hybridization amplitude are renormalized: $t_f\to b^2t_f$ and
$V_{df}\to b V_{df}$. The mean-field theory is carried out, enforcing
the constraint $b^{2}+ \langle n_{f}\rangle  = 1 $ on the average.
In addition, the chemical
potentials $\epsilon_d$ and $\varepsilon_f$ for both $d$-electrons and
$f$-holes are adjusted self-consistently to produce a band insulator,
$n_d+n_f=4$. This condition guarantees that four out of eight
doubly degenerate bands will be fully occupied. The details of our
mean-field calculation are given in the Supplementary Materials
section. Here we provide the final results of our calculations.

In Fig. \ref{Fig3} we show that the magnitude $b$
reduces with temperature, corresponding to a gradual
rise in the Sm valence, due to the weaker renormalization of the
$f$-electron level. The degree
 of mixed valence of Sm$^+$
is given then by $v=3-\langle n_{f}\rangle$. In our simplified
mean-field calculation, the smooth temperature
cross-over from Kondo insulating behavior to local moment metal at
high temperatures is crudely approximated by an abrupt second-order phase
transition.


\begin{figure}[h]
\begin{center}
\includegraphics[width=0.45\textwidth]{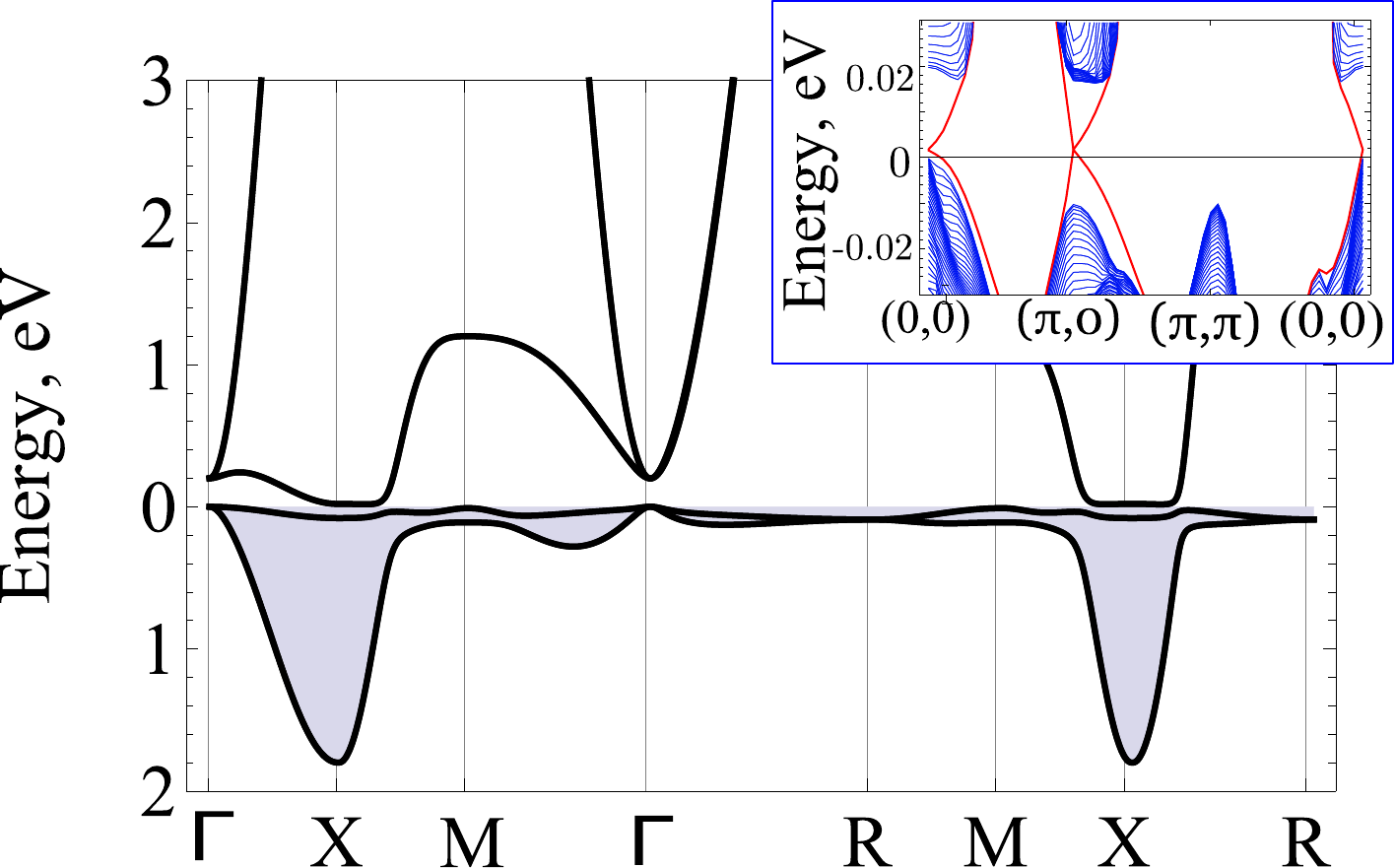}
\end{center}
\caption{Band structure consistent with PES and LDA studies of \smb{}
computed with the following
parameters: $n_f=0.48$
(or $b=0.73$), $V=0.05$ eV,$t_d = 2$ eV, $\mu_d=0.2$ eV , $\eta=\eta'=
-.3$, $\epsilon_f=-0.01$ eV ($\epsilon_{f0}=-0.17$ eV), $t_f=-.05$ eV,
$T=10^{-4}$ eV and the gap is $\Delta = 12$ meV. Shaded region
denotes filled bands. Inset shows the ground-state energy
computed for a slab of 80 layers
to illustrate the three gapless surface Dirac excitations
at the symmetry points $\hat{\Gamma}, \hat{X}' and \hat{X}''$.
 }
\label{Fig4}
\end{figure}

Fig. \ref{Fig4} shows the computed band structure for the cubic Kondo insulator
obtained from mean-field theory, showing the band
inversion between the $d$- and $f$-bands at the X points that generates
the strong topological insulator. Moreover,
as the value of the bare hybridization increases, there is a
maximum value beyond which the bands no longer invert and the Kondo
insulator becomes a conventional band insulator.

One of the interesting questions raised by this work concerns the many
body character of the Dirac electrons on the surface. Like the low-lying
excitations in the valence and conduction band, the surface states of
a TKI involve heavy quasiparticles of predominantly $f$-character.
The characteristic Fermi velocity of these excitations $v_{F}^{*}= Z
v_{F}$ is renormalized with respect to the conduction electron band
group velocities, where $Z= m/m^{*}$ is the mass renormalization of
the f-electrons.  In a band topological insulator, the penetration depth of
the surface states $\xi \sim v_{F}/\Delta $, where $\Delta $ is the band-gap,
scale that is significantly larger than a unit-cell
size. Paradoxically, even though the Fermi velocity of the Dirac cones
in a TKI is very low, we expect the characteristic
penetration depth $\xi$ of the heavy wavefunctions into the bulk to be of order
the lattice spacing $a$. To see this, we note that
$\xi \sim \frac{v_{F}^{*}}{\Delta_{g}}$, where the indirect gap of the Kondo
insulator  $\Delta_{g}$ is of order the Kondo temperature
$\Delta_{g}\sim T_{K}$. But  since $T_{K}\sim Z W$ , where $W$ is the
width of the conduction electron band, this implies that the
penetration depth of the surface excitations  $\xi \sim v_{F}/W\sim a$
is given by the size $a$ of the unit cell.   Physically, we can
interpret the surface Dirac cones as a result of broken Kondo singlets, whose spatial
extent is of order a lattice spacing. This feature is likely to make
the surface states rather robust against the purity of the bulk.

Various interesting questions are raised by our study.
Conventional Kondo insulators are most naturally  understood
as a
strong-coupling limit  of the Kondo lattice, where
local singlets form between a commensurate number of
conduction electrons and localized moments.
What then is the appropriate strong coupling description of
topological Kondo insulators, and can we understand the surface states
in terms of broken Kondo singlets?
A second question concerns the temperature dependence of the
hybridization gap. Experimentally, the hybridization gap observed in
Raman studies\cite{Sarrao1997} is seen to develop
in a fashion strongly reminiscent of the mean-field theory. Could this
indicate that fluctuations
about mean-field theory are weaker in a fully gapped Kondo lattice
than in its metallic counterpart?

We end with a few comments on the experimental consequences of the
above picture.   One of the most dramatic consequences of the quartet
model is the prediction of three Dirac
cones of surface excitation of substantially enhanced effective
mass. By contrast, were the underlying ground-state of the f-state a Kramers
doublet, then we would expect a single Dirac cone excitation.
These surface modes should be observable via various low energy
spectroscopies. For instance, the heavy mass of the quasi particles
should appear in Shubnikov-de Haas or
cyclotron resonance measurements.  The quasiparticle mass
 $m^* = \hbar  k_{F}/v_{F}^{*}\sim m_{e } (k_F a) Z^{-1 }$ will depend on
the Fermi momentum $k_F$ of the Dirac cones.
Scanning tunneling spectroscopy measurements of the
quasiparticle interference created by the Dirac conese,
and high resolution ARPES measurements may provide a direct way to
observe the predicted three Dirac cones.

To summarize, we have studied the cubic topological Kondo insulator,
incorporating the effect of a fourfold degenerate $f$-multiplet.
There are two main effects of the quartet states: first, they allow
the low fractional filling of the band required for strong topological
insulating behavior to occur in the almost integral valent environment
of the Kondo insulator; second, they double the degeneracy of the
band-states at the high-symmetry $\Gamma$ and $R$ points in the
Brillouin zone, removing these points from the calculation of the
Z$_{2}$ topological invariant so that the only important band
crossings occur at the the X or M points.  This leads to a prediction
that three heavy Dirac cones will form on the surfaces
\cite{Takimoto2011,Dai2012}.

We would like to thank A. Ramires, V. Galitski, K. Sun, S. Artyukhin  and
J. P. Paglione for stimulating discussions related to this work.
This work was supported by the Ohio Board of Regents Research
Incentive Program grant OBR-RIP-220573 (M.D.), DOE grant
DE-FG02-99ER45790 (V. A \& P. C.),
the U.S. National Science Foundation I2CAM International Materials
Institute Award, Grant DMR-0844115.

\twocolumngrid


\onecolumngrid

\newpage
\title{Supplementary Materials for Cubic Topological Kondo Insulators}

\author{Victor Alexandrov$^{1}$, Maxim Dzero$^{2}$ and Piers Coleman$^{1,3}$}

\affiliation{$^{1}$Center for Materials Theory, Department of Physics and Astronomy,
Rutgers University, Piscataway, NJ 08854-8019, USA}
\affiliation{$^{2}$ Department of Physics, Kent State University,
Kent, OH 44242, USA}
\affiliation{$^{3}$ Department of Physics, Royal Holloway, University
of London, Egham, Surrey TW20 0EX, UK.}
\date{\today}

%
\maketitle
%


\section{Supplementary materials for Cubic Topological Kondo insulators}

These notes provide:

\begin{itemize}
\item details of the derivation of the tight-binding
Hamiltonian for a cubic Kondo insulator.
\item derivation of the mean-field theory for the infinite $U$ limit
\item derivation of the mean-field equations.
\end{itemize}

\subsection{Rotation matrices}
To construct the Hamiltonian, we evaluate the hopping matrices along
the z-axis, and then carry out a unitary transformation to evaluate
the corresponding quantities for hopping along the x and y axes.

Consider a general rotation operator
\begin{equation}\label{rot oper}
    \mathcal{R} = R \Lambda,
\end{equation}
where $R$  and $ \Lambda$ describe $\pi/ 2$ rotations about a principle axis
of the crystal in real and spin space
respectively. The  Hamiltonian in a cubic environment is invariant under these
transformations: $H =
\mathcal{R}^\dag H \mathcal{R}$. We now write the directional dependence of
the Hamiltonian explicitly.
\begin{equation}\label{}
    H(x, y, z) =  (R \Lambda)^\dag   H(x, y, z) R \Lambda =  \Lambda^\dag  (R^\dag H(x, y, z) R) \Lambda.
\end{equation}
We can always choose the rotation $\mathcal{R}$ to transform in the cyclic manner: $x\rarrow y \rarrow z \rarrow x$, then substituting $z=y=0$,
 \begin{equation}\label{}
    H(x, 0, 0) =  \Lambda^\dag  (R_x^\dag H(x, 0, 0) R_x) \Lambda=  \Lambda_x^\dag   H(0,0, x)  \Lambda,
\end{equation}
hence we can assume the 3D Hamiltonian of the following form
\begin{equation}\label{1d 3d derived}
    H_{3D} = H(k_z) + \Lambda H_{}(k_x) \Lambda^\dag +\Lambda^\dag H_{}(k_y) \Lambda,
\end{equation}
 Where to rotate in the opposite direction we use $\Lambda^{-1}= \Lambda^\dag$. Using the Wigner D-functions we can construct the rotation in angular momentum space
\begin{equation}\label{rot oper 2}
    \Lambda=e^{-i \alpha J_z}e^{-i \beta J_y}e^{-i \gamma J_z} .
\end{equation}
Here, $\Lambda$ denotes the rotation operator
in the  ''ZYZ'' convention, corresponding to a rotation
around the $z$-axis, followed by a rotation around the $y$-axis, and
then the new -
$z$-axis. The matrix elements of this operator are then
\begin{eqnarray}\nonumber
    \begin{array}{lcl}
    D^{J}_{m' m} &=& \langle Jm'| \Lambda | Jm\rangle\cr
&=& [(j+m)!(j-m)!(j+m')!(j-m')!]^{1/2}
\cr
&&\times
{\sum}_\chi
\frac{(-1)^{\chi}}{(j-m'-\chi)!(j+m-\chi)!(\chi+m'-m)!\chi!}
\\
&&\times
\left(\cos\frac{\beta}{2}\right)^{2j+m-m'-2\chi}\left(-\sin\frac{\beta}{2}\right)^{m'-m+2\chi}
\cr
&& \times e^{-im'\alpha -i m\gamma}
\end{array}
\end{eqnarray}
so that upon rotating the state $ |Jm\rangle$ using $R_x$ we
obtain
$   \Lambda^{(j=5/2)}=D^{5/2}_{m m'} (0,\pi/2,\pi/2)$, and $
\Lambda^{(j=2)}=D^{2}_{m m'}\otimes D^{1/2}_{m m'}
(0,\pi/2,\pi/2)$. One can now obtain the transformation matrices for a
given multiplet ($e_g, t_{2g}, \Gamma_8 $ etc). For the $e_g$ doublet
we are to read off the matrices $M_{ij}=\langle i\vert
\Lambda_{x}\vert j\rangle $
from the transformation equation
\begin{equation}\label{rotation of the multiplet 1}
\Lambda_x \vert e_{g}:m\rangle  = \vert e_{g}:n\rangle M^d_{nm},
\end{equation}
where
  $\vert e_{g}:m\rangle  \equiv
\left\{ \vert d_{x^{2}-y^{2}} \uparrow\rangle,
\vert  d_{z^{2}}\uparrow\rangle ,
\vert d_{x^{2}-y^{2}} \downarrow\rangle,
\vert  d_{z^{2}}\downarrow\rangle
\right\}
$ is an $e_g$ doublet.
Similarly  for the $\Gamma_8$ quartet,
\begin{equation}\label{rotation of the multiplet 2}
\Lambda_x \vert \Gamma_{8},\alpha \rangle  = \vert
\Gamma_{8},\beta\rangle
M^f_{\beta\alpha },
\end{equation}
where the quartet is denoted by $\vert \Gamma_{8},\alpha\rangle $,
where $\alpha\in [1,4]$.
The result is,
\begin{equation}\label{rotation of the multiplet 3}
    M_{mn}^d=e^{\frac{i 3 \pi }{4}}
    \left(
\begin{array}{cccc}
 \frac{i}{2 \sqrt{2}} & \frac{1}{2} i \sqrt{\frac{3}{2}} & -\frac{i}{2 \sqrt{2}} & -\frac{1}{2} i \sqrt{\frac{3}{2}} \\
 -\frac{1}{2} i \sqrt{\frac{3}{2}} & \frac{i}{2 \sqrt{2}} & \frac{1}{2} i \sqrt{\frac{3}{2}} & -\frac{i}{2 \sqrt{2}} \\
 \frac{1}{2 \sqrt{2}} & \frac{\sqrt{\frac{3}{2}}}{2} & \frac{1}{2 \sqrt{2}} & \frac{\sqrt{\frac{3}{2}}}{2} \\
 -\frac{\sqrt{\frac{3}{2}}}{2} & \frac{1}{2 \sqrt{2}} & -\frac{\sqrt{\frac{3}{2}}}{2} & \frac{1}{2 \sqrt{2}} \\
\end{array}
\right),
\quad    M_{\alpha\beta}^f=e^{\frac{i 3 \pi }{4}}
    \left(
\begin{array}{cccc}
 \frac{1}{2 \sqrt{2}} & \frac{\sqrt{\frac{3}{2}}}{2} & -\frac{1}{2 \sqrt{2}} & -\frac{\sqrt{\frac{3}{2}}}{2} \\
 -\frac{\sqrt{\frac{3}{2}}}{2} & \frac{1}{2 \sqrt{2}} & \frac{\sqrt{\frac{3}{2}}}{2} & -\frac{1}{2 \sqrt{2}} \\
 -\frac{i}{2 \sqrt{2}} & -\frac{1}{2} i \sqrt{\frac{3}{2}} & -\frac{i}{2 \sqrt{2}} & -\frac{1}{2} i \sqrt{\frac{3}{2}} \\
 \frac{1}{2} i \sqrt{\frac{3}{2}} & -\frac{i}{2 \sqrt{2}} & \frac{1}{2} i \sqrt{\frac{3}{2}} & -\frac{i}{2 \sqrt{2}} \\
\end{array}
\right).
\end{equation}
Note that the rotation cyclically exchanges $x\rarrow y \rarrow z \rarrow x$, thus applying it three times gives overall "$-1$" due to fermionic statistics.
$$M^3 = -1.$$
Finally, we can use (\ref{rotation of the multiplet 3}) to derive the
hamiltonian defined in (5) and (6) of the main paper.

\subsection{Details of the mean-field theory}
The full Hamiltonian for the problem also contains a term describing the local Hubbard interaction between the $f$-electrons:
\beg
H_{f}=U\sum\limits_{i}\sum\limits_{\alpha=1}^4\sum\limits_{\beta\not=\alpha}f_{i\alpha}\dg f_{i\alpha}f_{i\beta}\dg f_{i\beta}
\en
We consider the infinite $U$, where we can project out all
states with occupation number larger than one by replacing the bare
f-electron fields by Hubbard operators. We represent the Hubbard
operators using a
slave boson representation, as follows:
\beg
f_{i\alpha}\dg \to X_{\alpha 0} (i)= f_{i\alpha}\dg b_i, \quad f_{i\alpha}\to X_{0\alpha } (i)= b_i\dg f_{i\alpha},
\en
supplemented by a constraint of no more than one $f$-electron per each site ($U=\infty$):
\beg\label{constraint}
\sum\limits_{\alpha=1}^4f_{i\alpha}\dg f_{i\alpha}+b_i\dg b_i=1.
\en
The partition function corresponding to the model Hamiltonian $H=H_c+H_f+H_{hyb}$ above and with constraint condition (\ref{constraint}) reads :
\beg
Z=\int\limits_{-\pi/\beta}^{\pi/\beta}\frac{\beta d\lambda}{\pi}\int{\cal D}(b,b\dg,f,f\dg,c,c\dg)
\exp\left(-\int\limits_0^\beta L(\tau)d\tau\right),
\en
where the Lagrangian $L(\tau)$ is
\beg\label{Lag}
\begin{split}
L=&\sum\limits_{i}b_i\dg\frac{d}{d\tau}b_i+\sum\limits_{ij}\sum\limits_{\alpha,\beta=1}^{4}f_{i\alpha}\dg
\left[\delta_{ij}\delta_{\alpha\beta}\left(\frac{d}{d\tau}+\varepsilon_f\right)+b_i t_{ij,\alpha\beta}^{(f)}b_j\dg\right] f_{j\beta}+\sum\limits_{\bk\sigma}\sum\limits_{a,b=1}^{2}c_{a\bk\sigma}\dg\left(\frac{d}{d\tau}+\varepsilon_{ab}^{(d)}(\bk)\right)c_{b\bk\sigma}\\&
+\frac{1}{2}\sum\limits_{\langle ij\rangle}\sum\limits_{\bk\sigma}\sum\limits_{a=1}^2\sum\limits_{\beta=1}^4\left(V_{ia\sigma,j\beta}
c_{ai\sigma}\dg b_{i}\dg f_{j\beta}+\textrm{h.c.}\right)+\sum\limits_{j}i\lambda_j\left(\sum\limits_{\alpha=1}^4f_{j\alpha}\dg f_{j\alpha}+b_j\dg b_j-1\right)
\end{split}
\en
Mean-field (saddle-point) approximation corresponds to the following values of the bosonic fields:
\beg\label{mf}
b_\bq(\tau)=b \delta_{\bq,0}, \quad i\lambda_\bq(\tau)=(E_f-\varepsilon_{f})\delta_{\bq,0},
\en
where both $a$ and $\varepsilon_f$ are $\tau$-independent.
In the mean-field theory  we choose a ``radial'' gauge where the
the phase of the $b$-field has been absorbed into the f-electron
fields.
Also, for an insulator, we need a filled quartet of states at each
site, so that
\beg\label{insulator}
n_c+n_f=4.
\en
Note, the parameter $b$ also renormalizes the $f$-hopping
elements. Indeed, it follows that when $i\neq j$,  $f_{i\alpha }\dg f_{j\beta }\to
X_{\alpha 0} (i)X_{0\beta} (j) = f_{i\alpha }\dg b_ib_j\dg f_{j\beta
}$. However, the onsite occupancy is unrenormalized by the slave boson
fields, since in the infinite $U$ limit, the onsite occupancy
$X_{\alpha \alpha } (i)= f\dg_{i\alpha } f_{i\alpha }$.

The first three terms together with the last term in the Lagrangian (\ref{Lag}) can be written using the new fermionic basis and (\ref{mf}). It follows:
\beg
\begin{split}
L_0(\tau)=
(b^2-1)(E_f-\varepsilon_f)+\sum\limits_{\bk\alpha}\sum\limits_{n=1}^{2}\tilde{f}_{n\bk\alpha}\dg
\left(\frac{d}{d\tau}+E_{n\bk}^{(f)}\right)\tilde{f}_{n\bk\alpha}+\sum\limits_{\bk\sigma}\sum\limits_{a=1}^{2}d_{a\bk\sigma}\dg\left(\frac{d}{d\tau}+
E_{a\bk}^{(d)}\right)d_{a\bk\sigma}
\end{split}
\en
where the renormalized $f$-electron dispersion is
\beg
E_{n\bk}^{(f)}=E_f+2t_fb^2\left(c_x+c_y+c_z+(-1)^n\sqrt{c_x^2+c_y^2+c_z^2-c_xc_y-c_yc_z-c_zc_x}\right), \quad n=1,2.
\en
Correspondingly, the hybridization matrix in the new fermionic basis can be obtained via unitary transformation with the matrix
\beg
{\cal U}_\bk=\left(\begin{matrix}
u_\bk & 0 & v_\bk & 0 \\
0 & u_\bk & 0 & v_\bk \\
-v_\bk & 0 & u_\bk & 0 \\
0 & -v_\bk & 0 & u_\bk
\end{matrix}
\right), \quad {\cal U}_\bk^{-1}=\left(\begin{matrix}
u_\bk & 0 & -v_\bk & 0 \\
0 & u_\bk & 0 & -v_\bk \\
v_\bk & 0 & u_\bk & 0 \\
0 & v_\bk & 0 & u_\bk
\end{matrix}
\right).
\en
Thus we need to calculate the elements of the matrix
\beg\label{HV}
\widetilde{H}_{V}=iV_{df}\frac{b}{2}\sum\limits_\bk\hat{d}_\bk\dg{\cal U}_\bk^{-1}H_{hyb}(\bk){\cal U}_\bk\tilde{f}_\bk+\textrm{h.c.}
\en
We can use the following relations
\beg
u_\bk^2-v_\bk^2=\frac{c_x+c_y-2c_z}{2R_\bk}, \quad u_\bk v_\bk=\frac{\sqrt{3}}{4R_\bk}(c_x-c_y).
\en
In our subsequent discussion it will be convenient to write $\widetilde{H}_{V}(\bk)$ in a more compact form. To derive the corresponding
expression we first recall that $H_{hyb}$ can be written as follows
\beg
\begin{split}
H_{hyb}(\bk)&=iV_{df}\frac{b}{2}\left[\hat{\phi}^x(\bk)\otimes\hat\sigma_x+\hat{\phi}^y(\bk)\otimes\hat\sigma_y+\hat{\phi}^z(\bk)\otimes\hat\sigma_z\right], \quad
\hat{\phi}_\alpha=\phi_0^\alpha\hat\tau_0+\phi_1^\alpha\hat\tau_z+\phi_2^\alpha\hat\tau_x, \\
{\vec \phi}_0&=\frac{2}{3}(\sin k_x,-\sin k_y,  \sin k_z), \quad {\vec \phi}_1=\frac{1}{3}(\sin k_x,-\sin k_y, -2\sin k_z),
\quad {\vec \phi}_2=+\frac{1}{\sqrt{3}}(\sin k_x,\sin k_y, 0).
\end{split}
\en
Again, note the change from minus sign to plus sign in front of $\vec{\phi}_2$ term. This yields the agreement
with the hybridization Hamiltonian obtained from the rotations method.
We also have
\beg
\hat{\cal U}_\bk=u_\bk\hat{\tau}_0\otimes\hat{\sigma}_0+iv_\bk\hat{\tau}_y\otimes\hat{\sigma}_0, \quad
\hat{\cal U}_\bk^{-1}=u_\bk\hat{\tau}_0\otimes\hat{\sigma}_0-iv_\bk\hat{\tau}_y\otimes\hat{\sigma}_0,
\en
After some algebra we find
\beg
\begin{split}
&\hat{\cal U}_\bk^{-1}\left(\hat\tau_0\otimes\hat\sigma_x\right)\hat{\cal U}_\bk=\hat{\tau}_0\otimes\hat\sigma_x, \\
&\hat{\cal U}_\bk^{-1}\left(\hat\tau_z\otimes\hat\sigma_x\right)\hat{\cal U}_\bk=(u_\bk^2-v_\bk^2)(\hat{\tau}_z\otimes\hat\sigma_x)
+2u_\bk v_\bk(\hat{\tau}_x\otimes\hat\sigma_x), \\
&\hat{\cal U}_\bk^{-1}\left(\hat\tau_x\otimes\hat\sigma_x\right)\hat{\cal U}_\bk=(u_\bk^2-v_\bk^2)(\hat{\tau}_x\otimes\hat\sigma_x)
-2u_\bk v_\bk(\hat{\tau}_z\otimes\hat\sigma_x).
\end{split}
\en
These results become much more transparent if we express $u_\bk$ and $v_\bk$ in terms of the angle $\theta_\bk$:
\beg
u_\bk=\cos(\theta_\bk/2), \quad v_\bk=\sin(\theta_\bk/2).
\en
Then we see that
\beg
\hat{\cal U}_\bk^{-1}\left(\hat{\phi}^x(\bk)\otimes\hat\sigma_x\right)\hat{\cal U}_\bk=\hat{\Phi}^x(\bk)\otimes\hat\sigma_x,
\en
where now
\beg
\hat{\Phi}^x=\Phi_0^x\hat{\tau}_0+\Phi_1^x\hat\tau_z+\Phi_2^x\hat\tau_x, \quad \Phi_0^x=\phi_0^x, \quad
\left(\begin{matrix} \Phi_1^x \\ \Phi_2^x\end{matrix}\right)=
\left[\begin{matrix}\cos\theta_\bk & -\sin\theta_\bk \\
\sin\theta_\bk & \cos\theta_\bk \end{matrix}\right]
\left(
\begin{matrix}\phi_1^x \\ \phi_2^x\end{matrix}
\right)
\en
Similarly, we find
\beg
\begin{split}
&\hat{\cal U}_\bk^{-1}\left(\hat{\phi}^y(\bk)\otimes\hat\sigma_y\right)\hat{\cal U}_\bk=\hat{\Phi}^y(\bk)\otimes\hat\sigma_y,
\quad \hat{\cal U}_\bk^{-1}\left(\hat{\phi}^z(\bk)\otimes\hat\sigma_z\right)\hat{\cal U}_\bk=\hat{\Phi}^z(\bk)\otimes\hat\sigma_z, \\
&\hat{\Phi}^y=\Phi_0^y\hat{\tau}_0+\Phi_1^y\hat\tau_z+\Phi_2^y\hat\tau_x, \quad \Phi_0^y=\phi_0^y, \quad
\left(\begin{matrix} \Phi_1^y \\ \Phi_2^y\end{matrix}\right)=
\left[\begin{matrix}\cos\theta_\bk & \sin\theta_\bk \\
-\sin\theta_\bk & \cos\theta_\bk \end{matrix}\right]
\left(
\begin{matrix}\phi_1^y \\ \phi_2^y\end{matrix}
\right), \\
&\hat{\Phi}^z=\Phi_0^z\hat{\tau}_0+\Phi_1^z\hat\tau_z+\Phi_2^z\hat\tau_x, \quad \Phi_0^z=\phi_0^z, \quad
\left(\begin{matrix} \Phi_1^z \\ \Phi_2^z\end{matrix}\right)=
\left[\begin{matrix}\cos\theta_\bk & -\sin\theta_\bk \\
\sin\theta_\bk & \cos\theta_\bk \end{matrix}\right]
\left(
\begin{matrix}\phi_1^z \\ \phi_2^z\end{matrix}
\right). \\
\end{split}
\en
Thus, we can re-write (\ref{HV}) as follows
\beg
\widetilde{H}_{V}=iV_{df}\frac{b}{2}\sum\limits_\bk\hat{d}_{\bk\alpha}\dg[\hat{\Phi}_\bk]_{\alpha\beta}\tilde{f}_{\bk\beta}+\textrm{h.c.},
\quad \hat{\Phi}_\bk=\hat{\Phi}^x(\bk)\otimes\hat\sigma_x+\hat{\Phi}^y(\bk)\otimes\hat\sigma_y+\hat{\Phi}^z(\bk)\otimes\hat\sigma_z
\en
Next we derive the mean-field equations.
\subsection{Derivation of the mean-field equations}
To derive the mean field equations we first integrate out $d$-electrons by making the following change of variables in the path integral:
\beg
\begin{split}
&\hat{d}_\bk\dg\to\hat{d}_\bk\dg+\hat{f}_\bk\dg\hat{\Phi}_\bk\dg\hat{G}_d(i\omega,\bk), \quad \hat{d}_\bk\to\hat{d}_\bk+\hat{G}_d(i\omega,\bk)\hat{\Phi}_\bk\hat{f}_\bk, \\
&\hat{G}_d^{-1}(i\omega,\bk)=\left(
\begin{matrix}
i\omega-E_{1\bk}^{(d)} & 0 & 0 & 0 \\
0 & i\omega-E_{1\bk}^{(d)} & 0 & 0 \\
0 & 0 & i\omega-E_{2\bk}^{(d)} & 0 \\
0 & 0 & 0 & i\omega-E_{2\bk}^{(d)}
\end{matrix}
\right).
\end{split}
\en
Then the resulting action is Gaussian and the $f$-electrons can be integrated out. This yields the effective action of the form
\beg
S_{eff}=(b^2-1)(E_f-\varepsilon_f)-T\sum\limits_{i\omega}\sum\limits_{\bk}\log\textrm{det}
\left[\hat{G}_{ff}^{-1}(i\omega,\bk)\right],
\quad \hat{G}_{ff}^{-1}=\hat{G}_f^{-1}(i\omega,\bk)-(V_{df}\frac{b}2)^2\hat{\Phi}_\bk\dg\hat{G}_d(i\omega,\bk)\hat{\Phi}_\bk.
\en
The renormalized $f$-electron correlation function $\hat{G}_{ff}^{-1}$ has a block diagonal form:
\beg
\hat{G}_{ff}^{-1}={(V_{df}b)^2\over 4}\left(\begin{matrix}
{\cal G}_{1f}^{-1}(i\omega,\bk){4 \over (V_{df}b)^2} & 0 &  -\frac{ \Delta_{1\bk}}{i\omega-E_{1\bk}^{(d)}}-\frac{ \Delta_{2\bk}}{i\omega-E_{2\bk}^{(d)}} & -\frac{ \widetilde{\Delta}_{1\bk}}{i\omega-E_{1\bk}^{(d)}}-\frac{ \widetilde{\Delta}_{2\bk}}{i\omega-E_{2\bk}^{(d)}}  \\
0 & {\cal G}_{1f}^{-1}(i\omega,\bk) {4 \over (V_{df}b)^2} & \frac{ \widetilde{\Delta}_{1\bk}^*}{i\omega-E_{1\bk}^{(d)}}+\frac{ \widetilde{\Delta}_{2\bk}^*}{i\omega-E_{2\bk}^{(d)}} & -\frac{ \Delta_{1\bk}^*}{i\omega-E_{1\bk}^{(d)}}-\frac{ \Delta_{2\bk}^*}{i\omega-E_{2\bk}^{(d)}} \\
-\frac{ \Delta_{1\bk}^*}{i\omega-E_{1\bk}^{(d)}}-\frac{ \Delta_{2\bk}^*}{i\omega-E_{2\bk}^{(d)}} &  \frac{ \widetilde{\Delta}_{1\bk}}{i\omega-E_{1\bk}^{(d)}}+\frac{ \widetilde{\Delta}_{2\bk}}{i\omega-E_{2\bk}^{(d)}} & {\cal G}_{2f}^{-1}(i\omega,\bk) {4 \over (V_{df}b)^2} & 0 \\
-\frac{ \widetilde{\Delta}_{1\bk}^*}{i\omega-E_{1\bk}^{(d)}}-\frac{ \widetilde{\Delta}_{2\bk}^*}{i\omega-E_{2\bk}^{(d)}} & -\frac{ \Delta_{1\bk}}{i\omega-E_{1\bk}^{(d)}}-\frac{ \Delta_{2\bk}}{i\omega-E_{2\bk}^{(d)}}  & 0 & {\cal G}_{2f}^{-1}(i\omega,\bk){4 \over (V_{df}b)^2}
\end{matrix}
\right),
\en
where the diagonal elements are given by
\beg
\begin{split}
&{\cal G}_{1f}^{-1}(i\omega,\bk)=i\omega-E_{1\bk}^{(f)}-(V_{df}\frac{b}{2})^2\left[\frac{\Phi_{1\bk}^{(+)}}{i\omega-E_{1\bk}^{(d)}}+
\frac{\Phi_{2\bk}}{i\omega-E_{2\bk}^{(d)}}\right], \\
&{\cal G}_{2f}^{-1}(i\omega,\bk)=i\omega-E_{2\bk}^{(f)}-(V_{df}\frac{b}{2})^2\left[\frac{\Phi_{2\bk}}{i\omega-E_{1\bk}^{(d)}}+
\frac{\Phi_{1\bk}^{(-)}}{i\omega-E_{2\bk}^{(d)}}\right],
\end{split}
\en
and we have introduced the following functions
\beg
\begin{split}
&\Phi_{1\bk}^{(\pm)}=(\Phi_{0}^x\pm\Phi_1^x)^2+(\Phi_{0}^y\pm\Phi_1^y)^2+(\Phi_{0}^z\pm\Phi_1^z)^2, \quad
\Phi_{2\bk}=(\Phi_{2}^x)^2+(\Phi_2^y)^2+(\Phi_2^z)^2, \\
&\Delta_{1\bk}=\Phi_2^z(\Phi_0^z+\Phi_1^z)+(\Phi_2^x+i\Phi_2^y)[\Phi_{0}^x+\Phi_1^x-i(\Phi_0^y+\Phi_1^y)], \\
&\Delta_{2\bk}=\Phi_2^z(\Phi_0^z-\Phi_1^z)+(\Phi_2^x-i\Phi_2^y)[\Phi_{0}^x-\Phi_1^x+i(\Phi_0^y-\Phi_1^y)], \\
&\widetilde{\Delta}_{1\bk}=(\Phi_2^x-i\Phi_2^y)(\Phi_0^z+\Phi_1^z)-\Phi_2^z[\Phi_{0}^x+\Phi_1^x-i(\Phi_0^y+\Phi_1^y)], \\
&\widetilde{\Delta}_{2\bk}=\Phi_2^z[\Phi_{0}^x-\Phi_1^x-i(\Phi_0^y-\Phi_1^y)]-(\Phi_2^x-i\Phi_2^y)(\Phi_0^z-\Phi_1^z).
\end{split}
\en
Then the determinant of the matrix $\hat{G}_{ff}$ is
\beg\label{Gff}
\begin{split}
\textrm{det}
\left[\hat{G}_{ff}^{-1}(i\omega,\bk)\right]=&\left\{{\cal G}_{1f}^{-1}(i\omega,\bk){\cal G}_{2f}^{-1}(i\omega,\bk)-(V_{df}\frac{b}{2})^4
\left(\frac{\Delta_{1\bk}}{i\omega-E_{1\bk}^{(d)}}+\frac{\Delta_{2\bk}}{i\omega-E_{2\bk}^{(d)}}\right)\left(\frac{\Delta_{1\bk}^*}{i\omega-E_{1\bk}^{(d)}}+\frac{\Delta_{2\bk}^*}{i\omega-E_{2\bk}^{(d)}}\right)\right.\\
&\left.-(V_{df}\frac{b}{2})^4
\left(\frac{\widetilde{\Delta}_{1\bk}}{i\omega-E_{1\bk}^{(d)}}+\frac{\widetilde{\Delta}_{2\bk}}{i\omega-E_{2\bk}^{(d)}}\right)\left(\frac{\widetilde{\Delta}_{1\bk}^*}{i\omega-E_{1\bk}^{(d)}}+\frac{\widetilde{\Delta}_{2\bk}^*}{i\omega-E_{2\bk}^{(d)}}\right)\right\}^2.
\end{split}
\en
Furthermore, we find that
\beg
|\Delta_{1\bk}|^2+|\widetilde{\Delta}_{1\bk}|^2=\Phi_{1\bk}^{(+)}\Phi_{2\bk}, \quad |\Delta_{2\bk}|^2+|\widetilde{\Delta}_{2\bk}|^2=\Phi_{1\bk}^{(-)}\Phi_{2\bk}
\en
so that certain terms in the expression (\ref{Gff}) will
cancel. Therefore, to derive the mean field equation we will need to
find the saddle point of the following effective action:
\beg\label{Seff_fin}
S_{eff}=(b^2-1)(E_f-\varepsilon_f)-2T\sum\limits_{i\omega}\sum\limits_{\bk}\log[P_{4\bk}(i\omega;E_f,b)], \quad
P_{4\bk}(i\omega)=\prod\limits_{i=1}^4(i\omega-\varepsilon_{i\bk}).
\en
where we have introduced the polynomial:
\beg
\begin{split}
&P_{4\bk}(i\omega)=(i\omega-E_{1\bk}^{(f)})(i\omega-E_{2\bk}^{(f)})(i\omega-E_{1\bk}^{(d)})(i\omega-E_{2\bk}^{(d)})+(V_{df}\frac{b}{2})^4\gamma_\bk^4-\\&-(V_{df}\frac{b}{2})^2\left\{
[\Phi_{2\bk}(i\omega-E_{1\bk}^{(f)})+\Phi_{1\bk}^{+}(i\omega-E_{2\bk}^{(f)})](i\omega-E_{2\bk}^{(d)})+
[\Phi_{2\bk}(i\omega-E_{2\bk}^{(f)})+\Phi_{1\bk}^{-}(i\omega-E_{1\bk}^{(f)})](i\omega-E_{1\bk}^{(d)})
\right\}, \\
&\gamma_\bk^4=\Phi_{1\bk}^{+}\Phi_{1\bk}^{-}+\Phi_{2\bk}^2-\Delta_{1\bk}\Delta_{2\bk}^*-\Delta_{1\bk}^*\Delta_{2\bk}-
\widetilde{\Delta}_{1\bk}\widetilde{\Delta}_{2\bk}^*-\widetilde{\Delta}_{1\bk}^*\widetilde{\Delta}_{2\bk}.
\end{split}
\en
Thus, two of the three mean field equations are formally given by
\beg
\frac{\partial S_{eff}}{\partial E_f}=0, \quad \frac{\partial S_{eff}}{\partial b}=0.
\en
We have
\beg
\begin{split}
&b^2-1+2\sum\limits_{i=1}^4\sum\limits_{\bk}\frac{f(\varepsilon_{i\bk})n_F(\varepsilon_{i\bk})}{\prod\limits_{l\not=i}(\varepsilon_{i\bk}-\varepsilon_{l\bk})}=0, \quad n_F(x)=\frac{1}{e^{\beta x}+1},
\end{split}
\en
where the function $f(\varepsilon)$ is:
\beg
\begin{split}
f(\lambda)&=2\lambda^3-\lambda^2[2(E_{1\bk}^{(d)}+E_{2\bk}^{(d)})+E_{1\bk}^{(f)}+E_{2\bk}^{(f)}]\\&+\lambda[2E_{1\bk}^{(d)}E_{2\bk}^{(d)}
+(E_{1\bk}^{(d)}+E_{2\bk}^{(d)})(E_{1\bk}^{(f)}+E_{2\bk}^{(f)})-(V_{df}\frac{b}{2})^2(\Phi_{1\bk}^{+}+\Phi_{1\bk}^{-}+2\Phi_{2\bk})]\\&+
(V_{df}\frac{b}{2})^2[(\Phi_{2\bk}+\Phi_{1\bk}^{-})E_{1\bk}^{(d)}+(\Phi_{2\bk}+\Phi_{1\bk}^{+})E_{2\bk}^{(d)}]-E_{1\bk}^{(d)}E_{2\bk}^{(d)}(E_{1\bk}^{(f)}+E_{2\bk}^{(f)})
\end{split}
\en
The derivation of the last mean-field equation can be compactly written as follows:
\beg
8(E_f-\varepsilon_f)+2\sum\limits_{i=1}^4\sum\limits_{\bk}\frac{\partial\varepsilon_{i\bk}}{\partial E_f}n_F(\varepsilon_{i\bk})=0.
\en


\end{document}